\documentclass[10pt]{article}
\usepackage{graphicx}
\graphicspath{{figures/}}

\def\Title#1{\begin{center} {\Large #1 } \end{center}}
\def\Author#1{\begin{center}{ \sc #1} \end{center}}
\def\Address#1{\begin{center}{ \it #1} \end{center}}

\newcommand\pubblock{\rightline{\begin{tabular}{l} Proceedings of the Fifth Annual LHCP\\ \pubnumber\\
         \pubdate  \end{tabular}}}

\newenvironment{Abstract}{\begin{quotation} \begin{center} 
             \large ABSTRACT \end{center}\bigskip 
      \begin{center}\begin{large}}{\end{large}\end{center} \end{quotation}}

\newenvironment{Presented}{\begin{quotation} \begin{center} 
             PRESENTED AT\end{center}\bigskip 
      \begin{center}\begin{large}}{\end{large}\end{center} \end{quotation}}

\def\beq{\begin{equation}}
\def\eeq#1{\label{#1}\end{equation}}
\def\eeqn{\end{equation}}

\def\beqa{\begin{eqnarray}}
\def\eeqa#1{\label{#1}\end{eqnarray}}
\def\eeqan{\end{eqnarray}}

\def\Dslash{\not{\hbox{\kern-4pt $D$}}}
\def\dslash{\not{\hbox{\kern-2pt $\del$}}}

\def\msb{{\bar{\ssstyle M \kern -1pt S}}}

\newcommand{\ttbar}{\ensuremath{t\bar{t}}}
\newcommand{\ifb}{\ensuremath{\mathrm{fb}^{-1}}}
\newcommand{\met}{\ensuremath{{E_{\mathrm{T}}^{\mathrm{miss}}}}}

\usepackage{color}
\usepackage{siunitx}
\usepackage{array} 
\usepackage{amssymb} %maths
\usepackage{amsmath} %maths
\usepackage{hyperref}
\usepackage{lineno}

\newcommand\pubnumber{ ATL-PHYS-PROC-2017-094 }
\newcommand\pubdate{\today}

\def\affiliation{
  On behalf of the ATLAS and CMS Collaborations\\
  School of Physics A28\\
  University of Sydney\\
  NSW 2006, Australia
}

\begin{document}
%\linenumbers
% large size for the first page
\large
\begin{titlepage}
\pubblock

%% Change the title, name, abstract
%% Title 
\vfill
\Title{  Single top-quark production via $tW$, $tZq$, and $s$-channel with ATLAS~and~CMS}
\vfill

%  if you need to add the support use this, fill the \support definition above. 
%   \Author{ FIRSTNAME LASTNAME \support }
\Author{ Kevin Finelli \footnote{Now at Bostun University} }
\Address{\affiliation}
\vfill
\begin{Abstract}

  Analyses of single top-quark production via $tW$ production, via $tZq$ production, and in the $s$-channel are presented, showing the scope of recent ATLAS and CMS analyses of $pp$ collisions produced at the LHC at $\sqrt{s}=7,8$, and $\SI{13}{TeV}$.
  Cross-section values and signal significance estimates are established, confirming good agreement with the most precise available fixed-order calculations for each production mode.

\end{Abstract}
\vfill

% DO NOT CHANGE 
\begin{Presented}
The Fifth Annual Conference\\
 on Large Hadron Collider Physics \\
Shanghai Jiao Tong University, Shanghai, China\\ 
May 15-20, 2017
\end{Presented}
\vfill
\end{titlepage}
\def\thefootnote{\fnsymbol{footnote}}
\setcounter{footnote}{0}
%

% normal size for the rest
\normalsize 

%% Your paper should be entered below. 

\section{Introduction}

\begin{figure}[htb]
  \begin{center}
    \begin{tabular}{ccc}
      \includegraphics[width=.25\textwidth,trim={0 10em 5em 10em},clip]{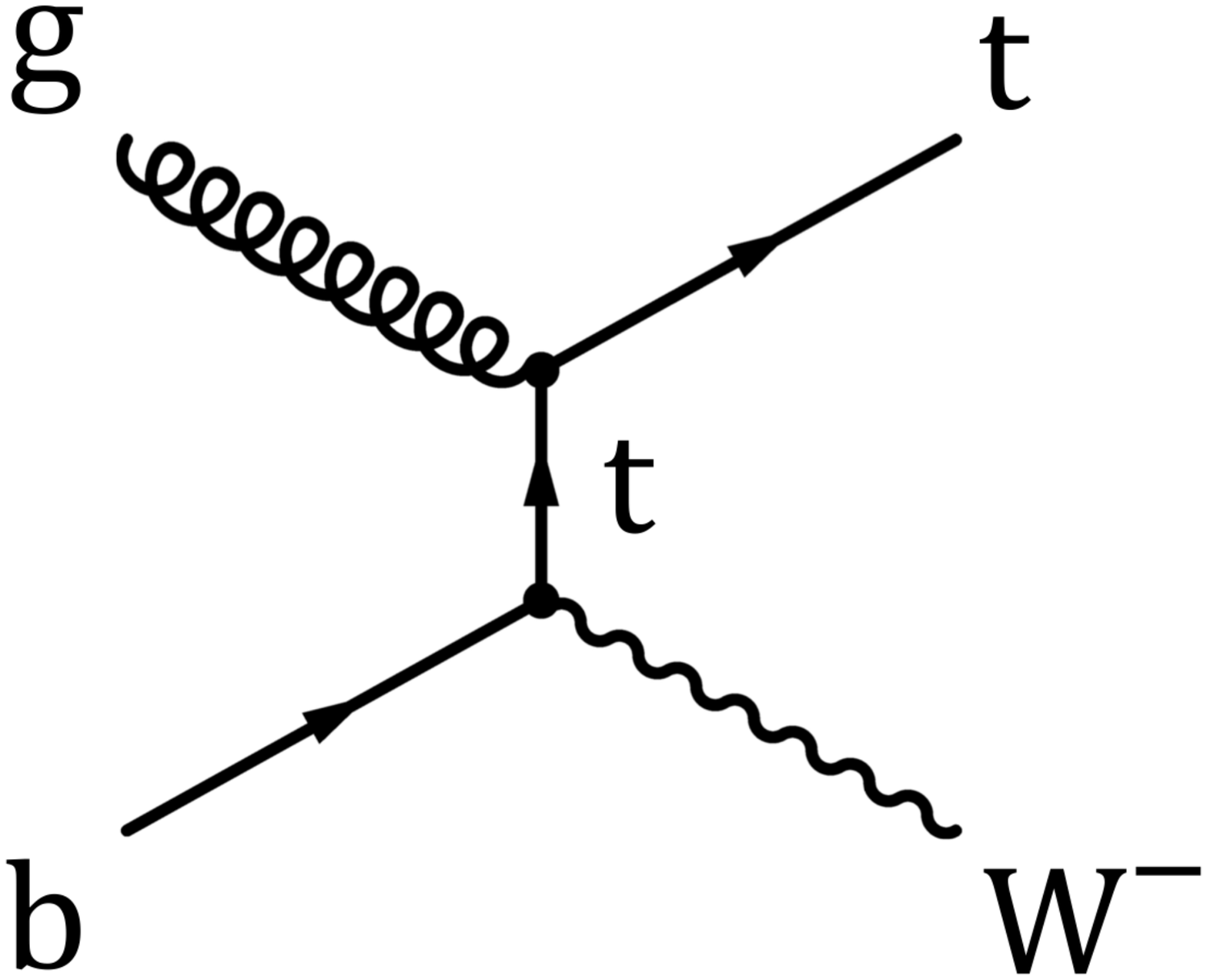} &
          \includegraphics[width=.35\textwidth,trim={7em 25em 22em 0em},clip]{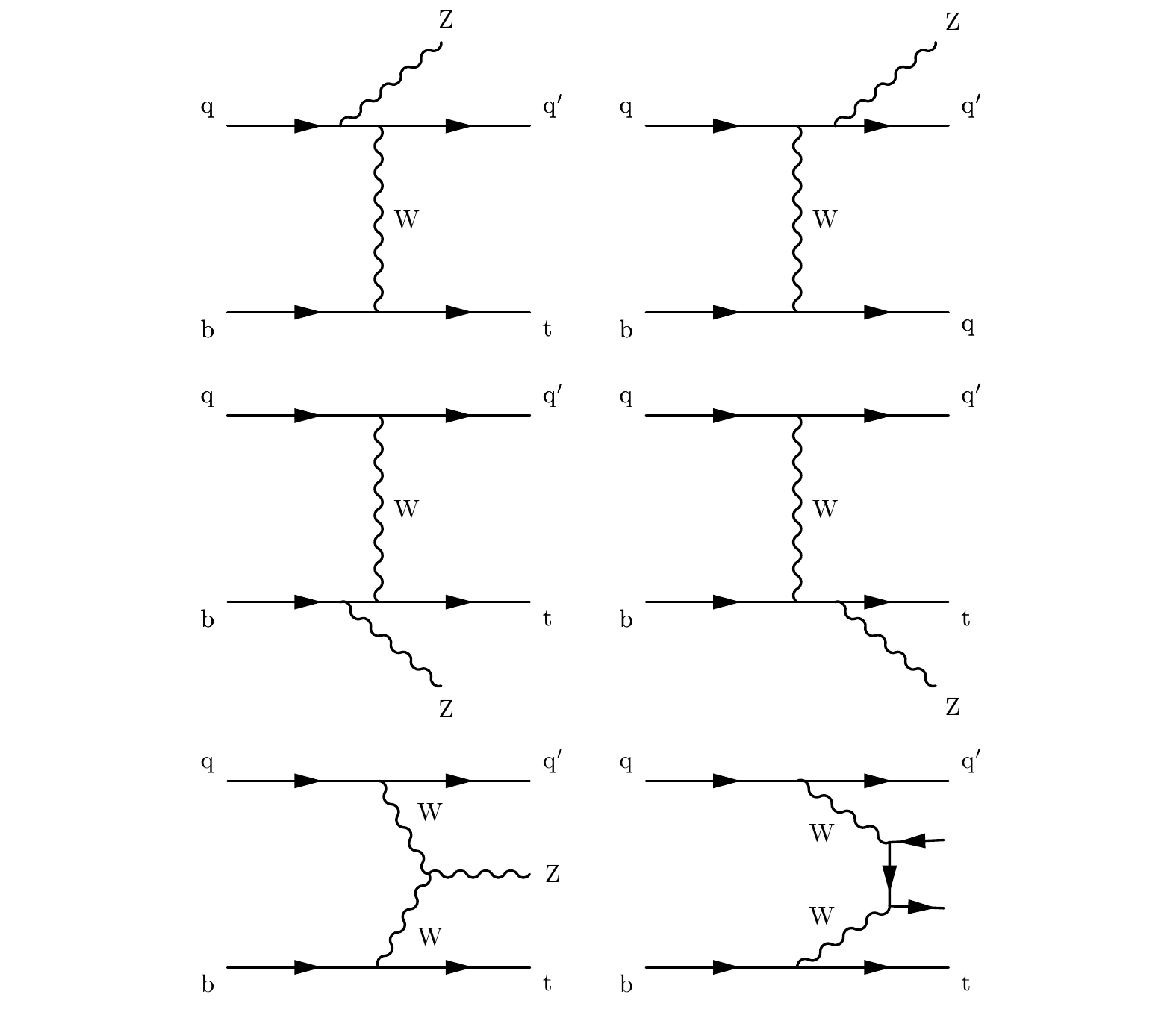} &
      \includegraphics[width=.25\textwidth,trim={4em 10em 5em 20em},clip]{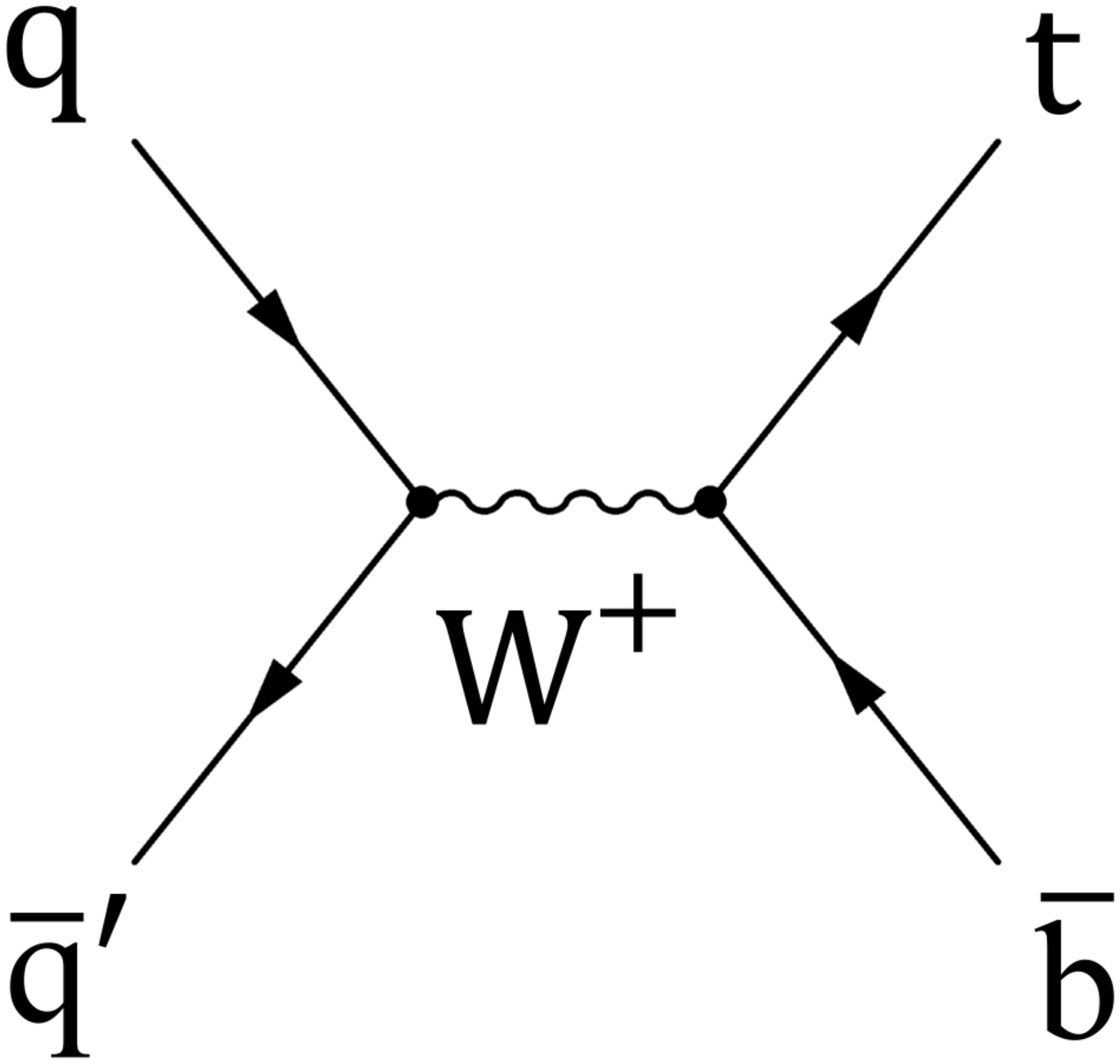} \\
          (a) & (b) & (c) \\
    \end{tabular}
  \end{center}
  \label{fig:feyn}
      \caption{ Example leading order diagrams contributing to single top-quark production via (b) $tW$, (c) $tZq$, and (c) $s$-channel. }
\end{figure}

The ATLAS~\cite{Aad:2008zzm} and CMS~\cite{Chatrchyan:2008aa} experiments have had a successful program in the study of the top quark using $pp$ collisions in the Large Hadron Collider~(LHC).
The top quark is the most massive known elementary particle, and its large mass makes it a uniquely important part of the Standard Model (SM), playing a role in vacuum stability as well as having the strongest Yukawa coupling to the Higgs boson of any particle.
Although the dominant production mode for the top quark at the LHC is pair production, production of a single top quark may also occur through $t$-channel, $tW$, $tZq$, $s$-channel, or other rarer processes.
Figure~\ref{fig:feyn} shows diagrams for $tW$, $tZq$, and $s$-channel production, which are the focus of this proceeding.
These rarer single top-quark production modes have distinct physics motivations from $t$-channel production, including possible beyond SM (BSM) resonances that decay to heavy quarks and may appear in $tb$ production in the $s$-channel. 
Also, modifications of the $tWb$ vertex may appear with different effective field theory operators in $tW$ production compared to other modes, motivating the study of this channel.

%The ATLAS and CMS measurements presented in these proceedings were performed on $pp$-collision data collected in LHC running from 2011 to 2015, at center-of-mass energies of $\sqrt{s}={7}\,\mathrm{TeV}$ (corresponding to $5.1\,\mathrm{fb}^{-1}$ for CMS in 2011) $\sqrt{s}={8}\,\mathrm{TeV}$ (corresponding to $20.2\,\mathrm{fb}^{-1}$ for ATLAS and $19.7\,\mathrm{fb}^{-1}$ for CMS in 2012) and $\sqrt{s}={13}\,\mathrm{TeV}$ (corresponding to $3.2\,\mathrm{fb}^{-1}$ for ATLAS in 2015).

%ATLAS 8 TeV tW
\section{ATLAS $tW$ cross-section at $\sqrt{s}=\SI{8}{\TeV}$}
\begin{figure}[htb]
  \begin{center}
    \begin{tabular}{ccc}
      \includegraphics[width=.3\textwidth]{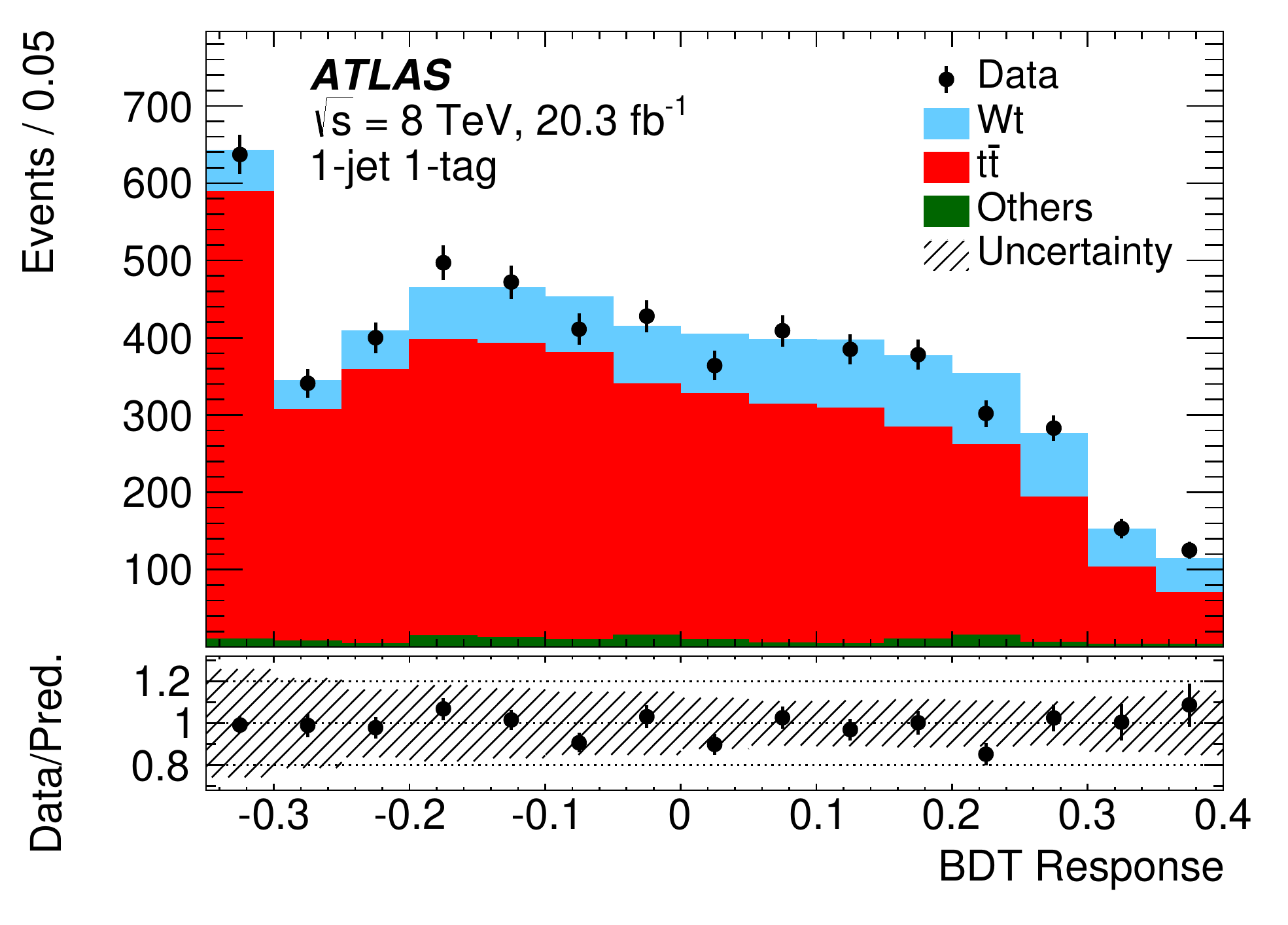} &
      \includegraphics[width=.3\textwidth]{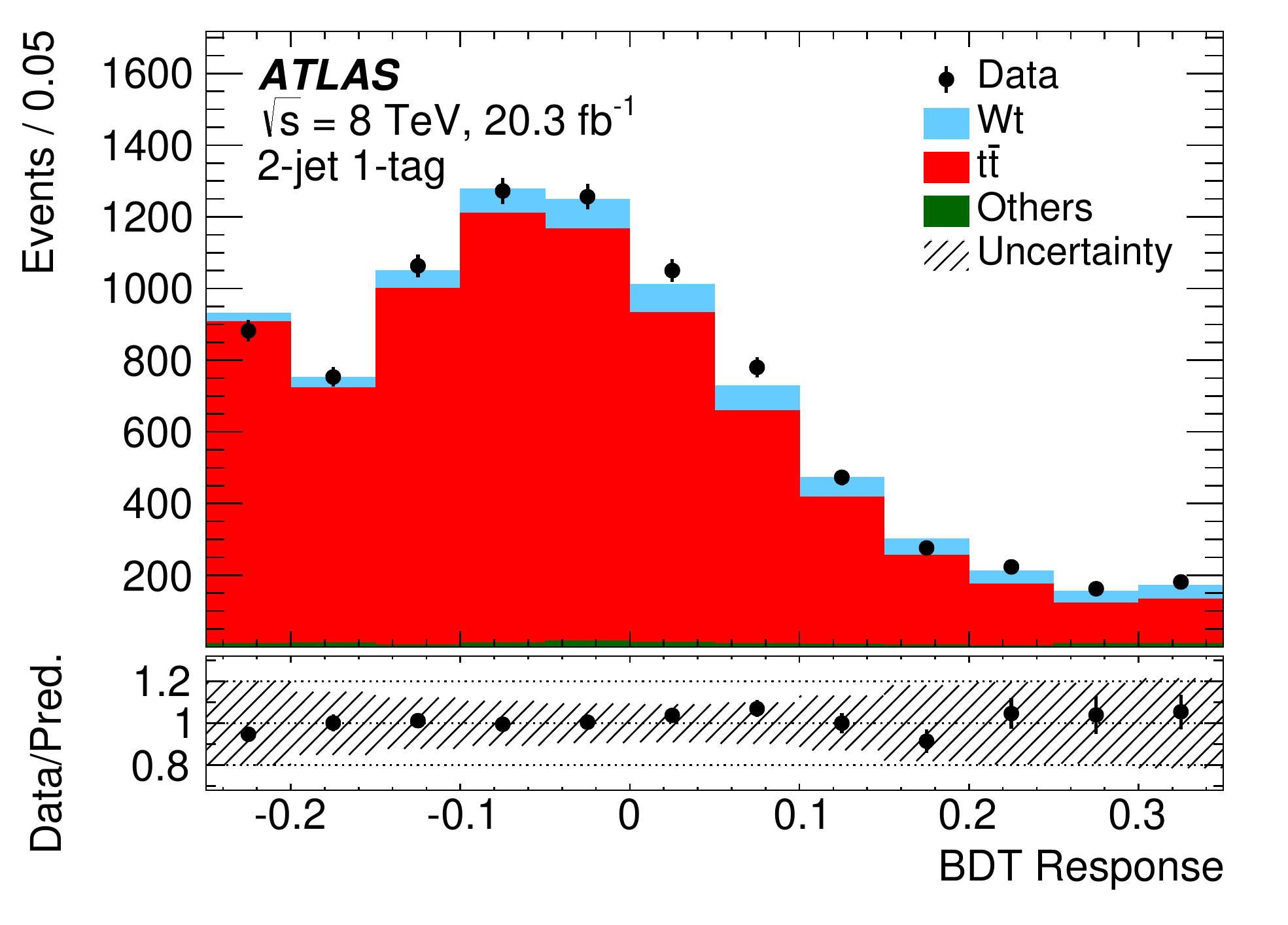} &
      \includegraphics[width=.3\textwidth]{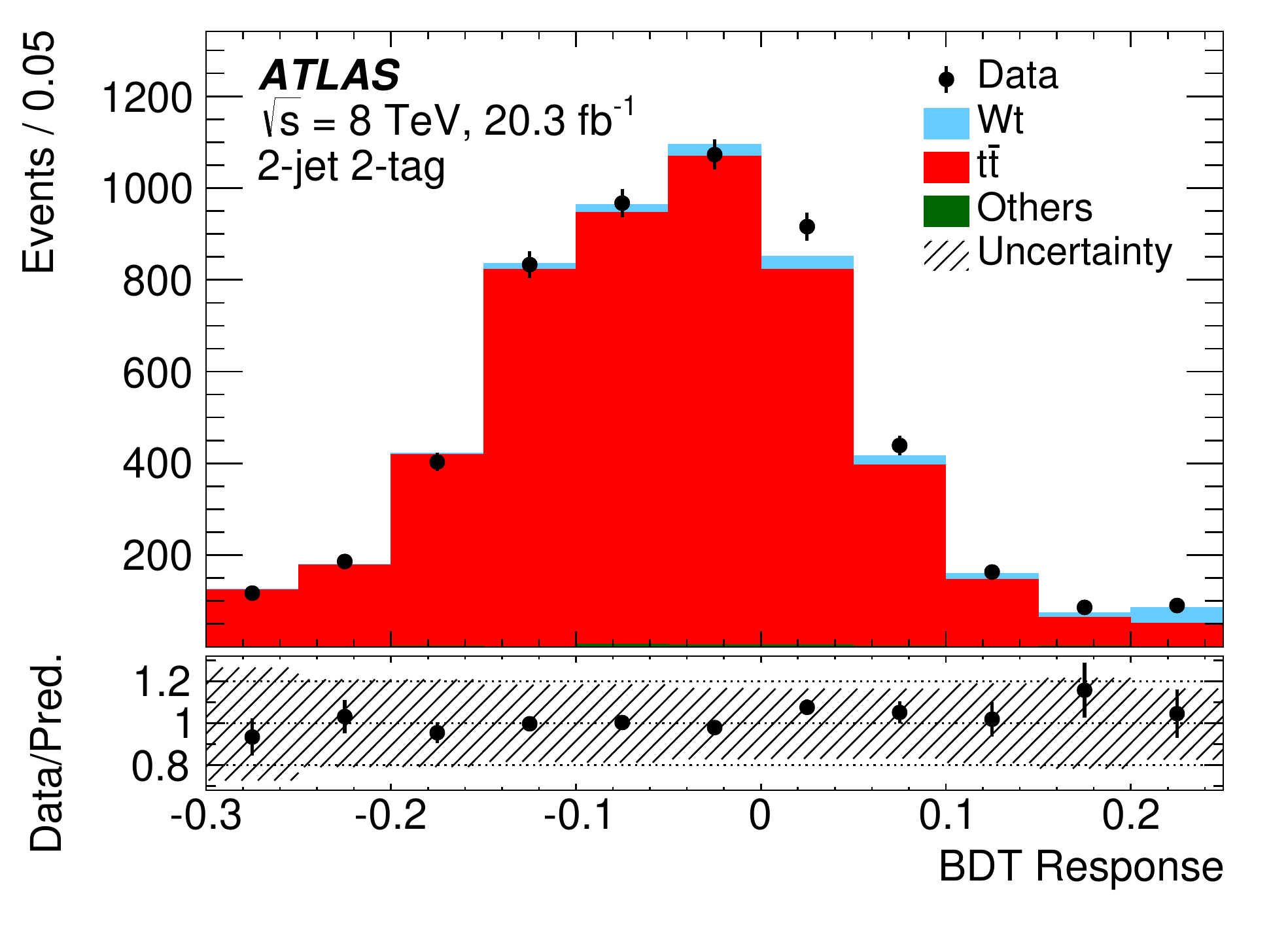} \\
      (a) & (b) & (c)\\
    \end{tabular}
  \end{center}
  \caption{BDT response for (a) 1-jet 1-tag, (b) 2-jet 1-tag and (c) 2-jet 2-tag events~\cite{Aad:2015eto}.
    Each contribution is normalized to its expectation.
    The hatched area represents the sum in quadrature of the statistical and systematic uncertainties.
    The first bin includes the underflow and the last bin the overflow.}
  \label{fig:atlas_tw8}
\end{figure}

%https://atlas.web.cern.ch/Atlas/GROUPS/PHYSICS/PAPERS/TOPQ-2012-20/
A cross-section measurement is performed with $\SI{20.3}{fb^{-1}}$ of ATLAS data at $\sqrt{s}=\SI{8}{\TeV}$~\cite{Aad:2015eto}.
Events are selected with two leptons ($e$ or $\mu$) and one or two hadronic jets, at least one of which are required to be tagged as likely to contain $B$ hadrons ($b$-tagged jets).
Three regions are defined based on the number of jets and $b$-tagged jets, a signal region (1j1b), and two \ttbar{} control regions (2j1b, 2j2b).
The $tW$ signal is separated from \ttbar{} background by employing a boosted decision tree (BDT)~\cite{BDT}.
Figure~\ref{fig:atlas_tw8} shows the output of BDT response for data as well as signal and background predictions.
A global maximum likelihood fit to the BDT response in each of the three regions is then performed to extract the cross-section and constrain uncertainties.

The analysis also employs a modified fit of $tW+\ttbar{}$ in only the 1j1b region to extract a fiducial cross-section of $WWb$ and $WWbb$ events defined by the selection of two leptons and exactly one $b$-tagged jet in the fiducial acceptance.
In the fiducial measurement the impact of theoretical modeling uncertainties is significantly smaller compared to full cross-section measurement.
The values of the measured full cross-section $\sigma_{\text{obs}}$ and fiducial cross-section $\sigma_{\text{obs}}^\text{fid}$ are:

\begin{eqnarray*}
  \sigma_{\text{obs}} =& 23.0 \pm 1.3\, \text{(stat)} ^{+3.2}_{-3.5}\, \text{(syst)} \pm 1.1\, \text{(lumi) pb}\\
  \sigma_{\text{obs}}^\text{fid} =& 0.85 \pm 0.01\,\text{(stat)}^{+0.06}_{-0.07}\,\text{(syst)}\pm 0.03\,\text{(lumi) pb}.
\end{eqnarray*}
The inclusive cross section is compared to the approximate NNLO prediction from Kidonakis~\cite{Kidonakis:2010ux}, $\sigma_{\text{appx. NNLO}} = 22.4 \pm 1.5\, \text{pb}$.
The data show a signal significance of $7.7 \sigma$, while $6.9\sigma$ significance is expected.
The systematic uncertainties with the largest impacts are jet- and \met-related uncertainties as well as signal modeling uncertainties related to initial- and final-state QCD radiation.
This ATLAS measurement is the most precise $tW$ cross-section measurement at $\sqrt{s}=\SI{8}{\TeV}$, and the first observation of the $tW$ process with significance greater than $5\sigma$.

%CMS 8 TeV tW
%http://cms-results.web.cern.ch/cms-results/public-results/publications/TOP-12-040/index.html
\section{CMS $tW$ cross-section at $\sqrt{s}=\SI{8}{\TeV}$}
\begin{figure}[htb]
  \begin{center}
    \begin{tabular}{ccc}
      \includegraphics[width=.3\textwidth]{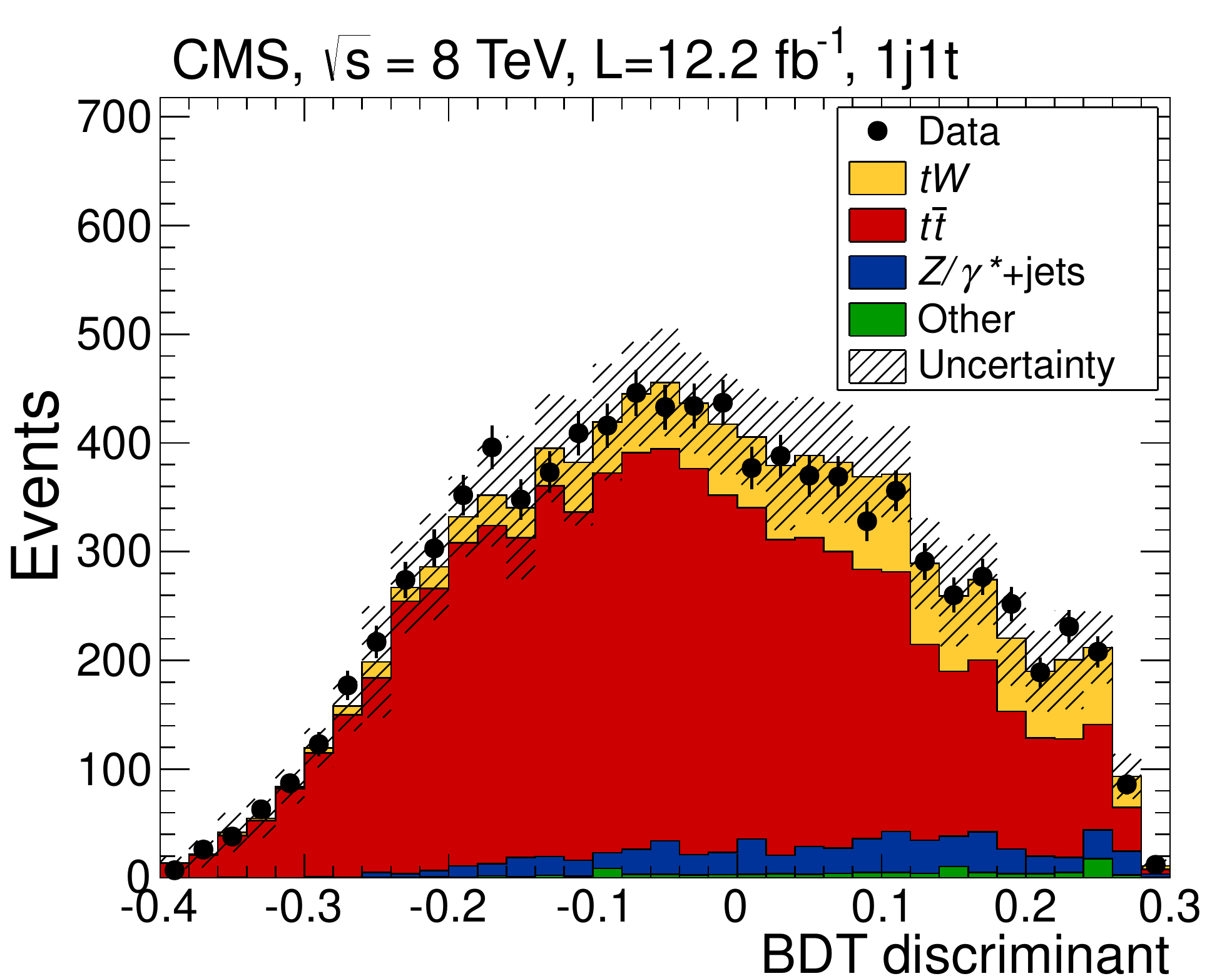} &
      \includegraphics[width=.3\textwidth]{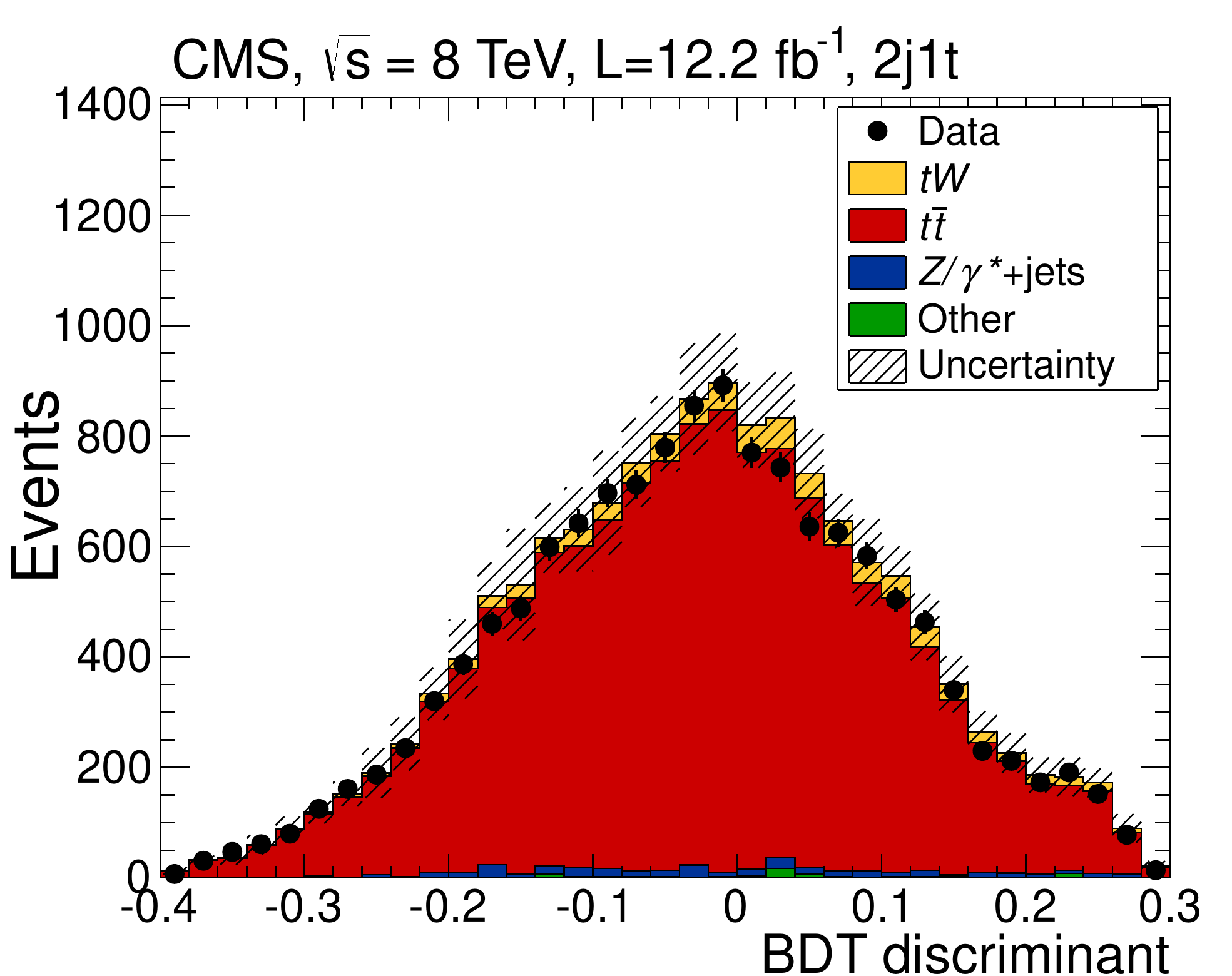} &
      \includegraphics[width=.3\textwidth]{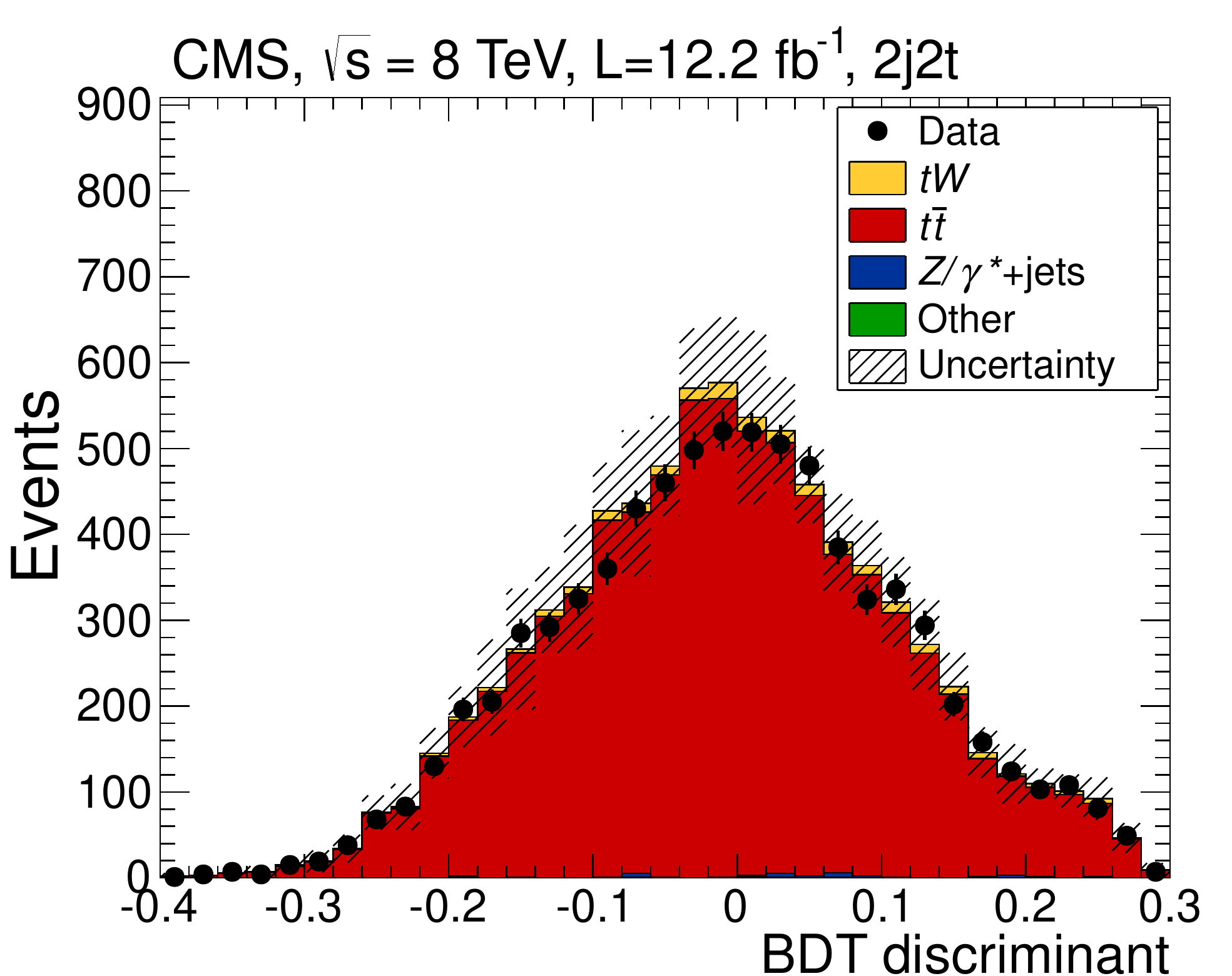} \\
      (a) & (b) & (c) \\
    \end{tabular}
  \end{center}
  \caption{The BDT discriminant, in the signal region (a) 1j1b and control regions (b) 2j1b and (c) 2j2b for all final states combined~\cite{Chatrchyan:2014tua}.
  Shown are data (points) and simulation (histogram).
  The hatched band represents the combined effect of all sources of systematic uncertainty.}
  \label{fig:cms_tw8}
\end{figure}

The cross-section for $tW$ production at $\sqrt{s}=\SI{8}{\TeV}$ is measured with $\SI{12.1}{fb^{-1}}$ of data by the CMS experiment~\cite{Chatrchyan:2014tua}.
Selected events are required to pass dilepton and \met{} criteria, then categorized as 1j1b, 2j1b, 2j2b in a similar manner as described above.
The $tW$ signal is separated from the \ttbar{} background with a BDT, shown in Figure~\ref{fig:cms_tw8}, and a global maximum likelihood fit is performed to extract the cross-section and constrain systematic uncertainties.
The most significant BDT input variables in terms of separation power are the number of loose jets, the number of central loose jets, and the number of $b$-tagged loose jets, followed by kinematics of leptons, jet, \met.
The cross-section is measured to be $\sigma_{\text{obs}} = 23.4 \pm 5.4\, \text{pb}$, which agrees well with the prediction of Ref.~\cite{Kidonakis:2010ux} shown above.
The observed signal significance is $6.1 \sigma$, with an expected significance of $5.4\sigma$.
Additionally, a cross-check cut-and-count analysis is performed, which finds results consistent with the main BDT and fit analysis.
The measurement uncertainty is systematics-dominated, and it arises primarily due to theoretical modeling uncertainties on $tW$ and \ttbar.

%ATLAS 13 TeV tW
%https://atlas.web.cern.ch/Atlas/GROUPS/PHYSICS/PAPERS/TOPQ-2015-16/
\section{ATLAS $tW$ cross-section at $\sqrt{s}=\SI{13}{\TeV}$}
\begin{figure}[hbt]
  \begin{center}
    \includegraphics[width=0.7\textwidth]{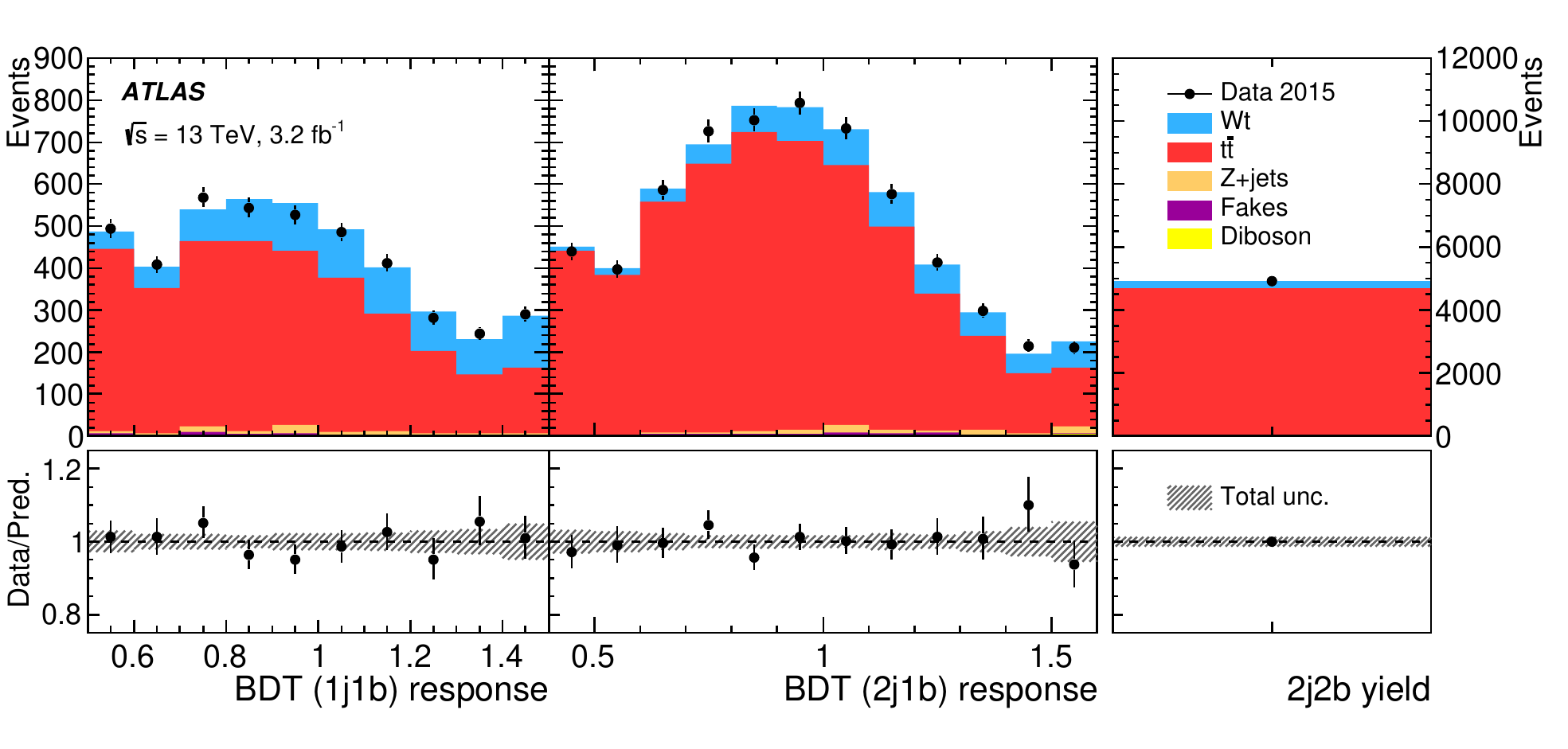}
  \end{center}
  \caption{Post-fit distributions in the signal and control regions 1j1b, 2j1b, and 2j2b~\cite{Aaboud:2016lpj}.
  The error bands represent the total uncertainties in the fitted results.
  The upper panels give the yields in number of events per bin, while the lower panels give the ratios of the numbers of observed events to the total prediction in each bin.}
  \label{fig:atlas_tw13}
\end{figure}

A measurement of the $tW$ production cross-section is performed by ATLAS using $\SI{3.2}{\ifb}$ of data at $\sqrt{s}=\SI{13}{\TeV}$~\cite{Aaboud:2016lpj}.
    The overall strategy is similar to Ref.~\cite{Aad:2015eto}; dilepton events are selected and separated into three regions: 1j1b, 2j1b, and 2j2b.
A BDT is used in the 1j1b and 2j1b regions, which have significant signal contributions, while a single bin is used for 2j2b where signal contribution is minimal, as shown in Figure~\ref{fig:atlas_tw13}.
A combined likelihood fit is performed in the three regions, which determines the $tW$ cross-section and constrains uncertainties.
The $tW$ cross-section is measured to be:

\begin{equation*}
  \sigma_{\text{obs}} = 94 \pm 10\,\text{(stat)} ^{+28}_{-22}\,\text{(syst)} \pm 2\,\text{(lumi) pb},
\end{equation*}
which is in good agreement with the approximate NNLO prediction:

\begin{equation*}
  \sigma_{\text{appx. NNLO}} = 71.7 \pm 3.9\, \text{pb}
\end{equation*}
from Ref.~\cite{Kidonakis:2010ux}.
The most significant systematic uncertainties come from the $tW$ parton shower generator, jet energy scale, and \ttbar{} initial- and final-state radiation generator tuning.
The likelihood fit is able to significantly constrain some large uncertainties, including the $tW$ parton shower generator, and the \ttbar{} initial- and final-state radiation tuning.
Future measurements using the full Run~2 dataset should be able to further constrain these uncertainties and produce a more precise result.

%%%%%%%%%%%%%%%%%%%%%%
%CMS tZq 8 TeV
%http://cms-results.web.cern.ch/cms-results/public-results/publications/TOP-12-039/index.html
\section{CMS $tZq$ search at $\sqrt{s}=\SI{8}{\TeV}$}
\begin{figure}[htb]
  \begin{center}
    \begin{tabular}{cc}
      \includegraphics[width=0.35\textwidth]{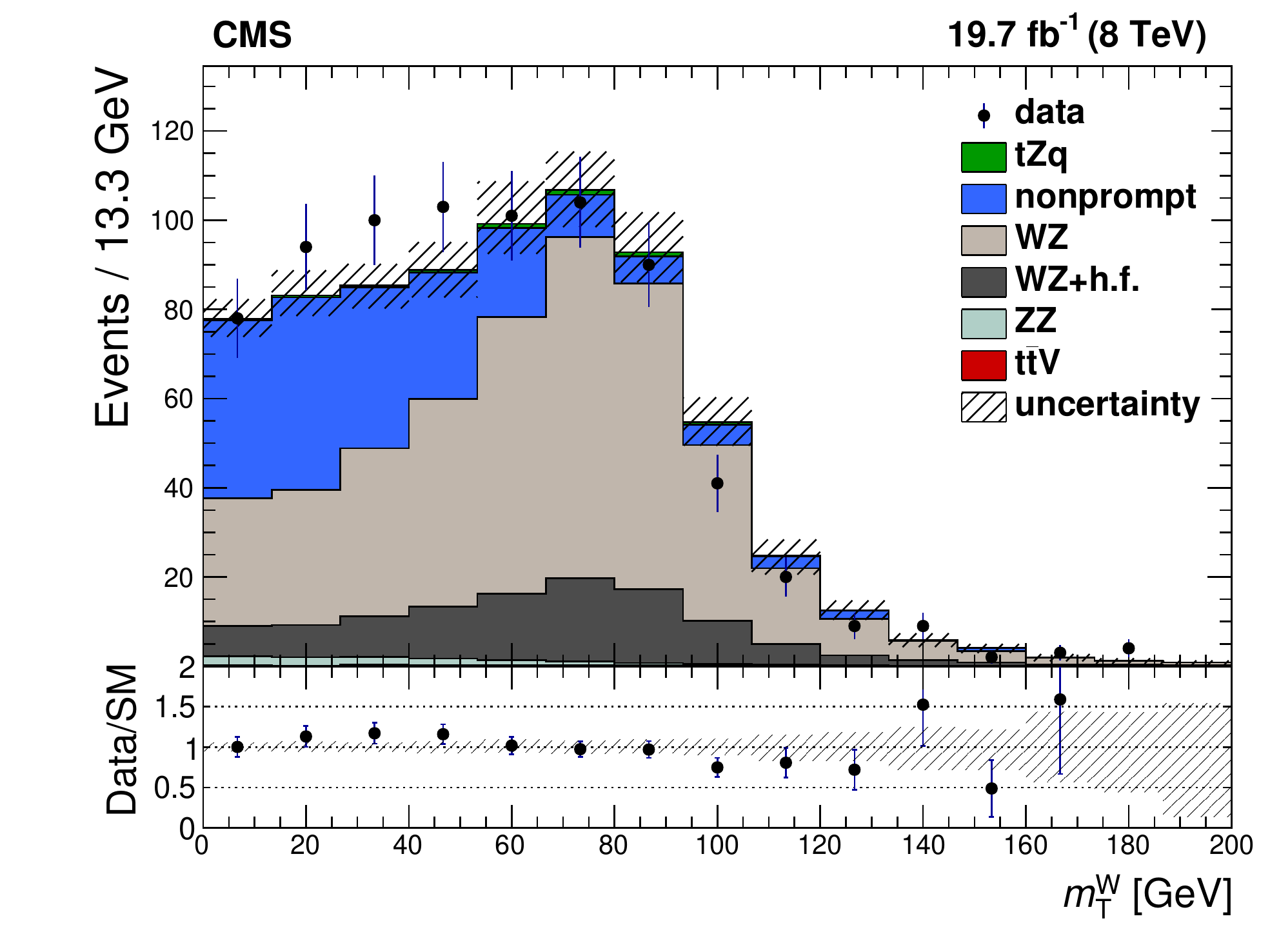} &
      \includegraphics[width=0.35\textwidth]{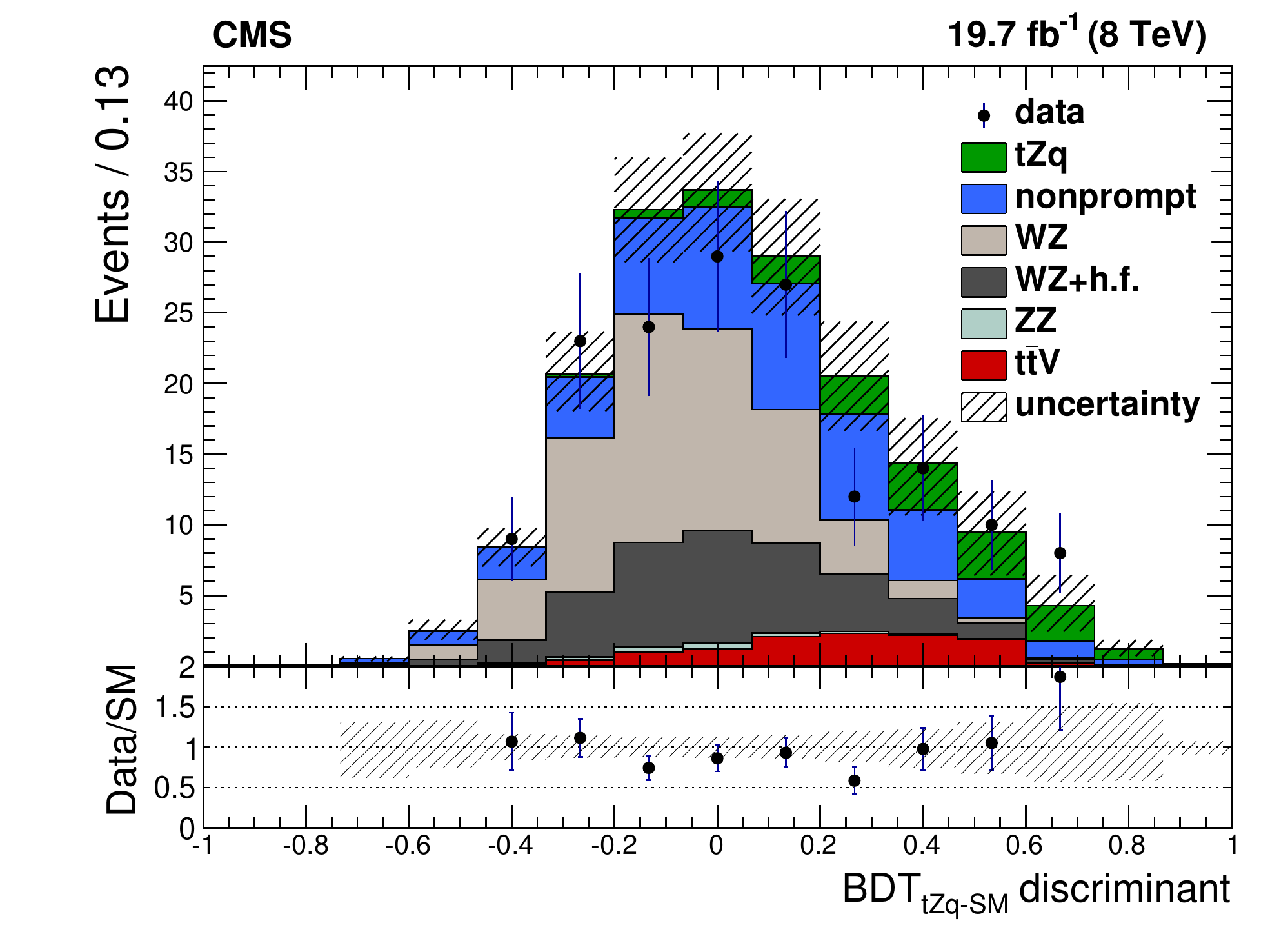} \\
      (a) & (b) \\
    \end{tabular}
  \end{center}
    \caption{Data-to-prediction comparison after performing the fit for (a) the $m^W_\mathrm{T}$ distribution in the control region and (b) the $\mathrm{BDT}_{tZq-\mathrm{SM}}$ responses in the signal region~\cite{Sirunyan:2017kkr}.
    The four lepton channels containing all combinations of three leptons are combined.
    The lower panel shows the ratio between observed and predicted yields, including the total uncertainty on the prediction.}
    \label{fig:cms_tzq}
\end{figure}

A search is performed for $tZq$ production with $\SI{19.7}{\ifb}$ of CMS $\sqrt{s}=\SI{8}{\TeV}$ data~\cite{Sirunyan:2017kkr}, where the $Z$ boson decays to two charged leptons and the $W$ from the top quark decays leptonically.
Events containing three leptons, at least two jets including at least one $b$-tagged jet are selected for analysis.
A BDT is trained to separate $tZq$ signal from the main backgrounds, $\ttbar Z$ and $WZ$.
A template fit is performed to data using the distributions of the BDT discriminant in the signal region and $m_{\text{T}}^{W}$ in a $b$-veto control region, the results of which are shown in Figure~\ref{fig:cms_tzq}.
The measured cross-section times branching ratio is:

\begin{equation*}
  \sigma_\text{obs}(pp\to t \ell^+ \ell^-)\mathcal{B}(t \to \ell \nu b) =  10^{+8}_{-7}\, \text{fb},
\end{equation*}
which is compared to the NLO calculation from \textsc{MG5\_aMC@NLO}~\cite{Alwall:2014hca}:

\begin{equation*}
  \sigma_\text{NLO}(pp\to t \ell^+ \ell^-)\mathcal{B}(t \to \ell \nu b) = 8.18^{+0.59}_{-0.03}\,\text{(scale)}\,\text{fb},
\end{equation*}
and found to be in good agreement within the uncertainties, which are driven mainly by limited data statistics.
The combined signal significance observed (expected) is $2.4\sigma\, (1.8\sigma)$.

%%%%%%%%%%%%%%%%%%%%%%
%CMS 7,8 TeV tb
%http://cms-results.web.cern.ch/cms-results/public-results/publications/TOP-13-009/index.html
\section{CMS $tb$ search at $\sqrt{s}=7,\SI{8}{\TeV}$}
\begin{figure}[htb]
  \begin{tabular}{ccc}
    \includegraphics[width=.3\textwidth]{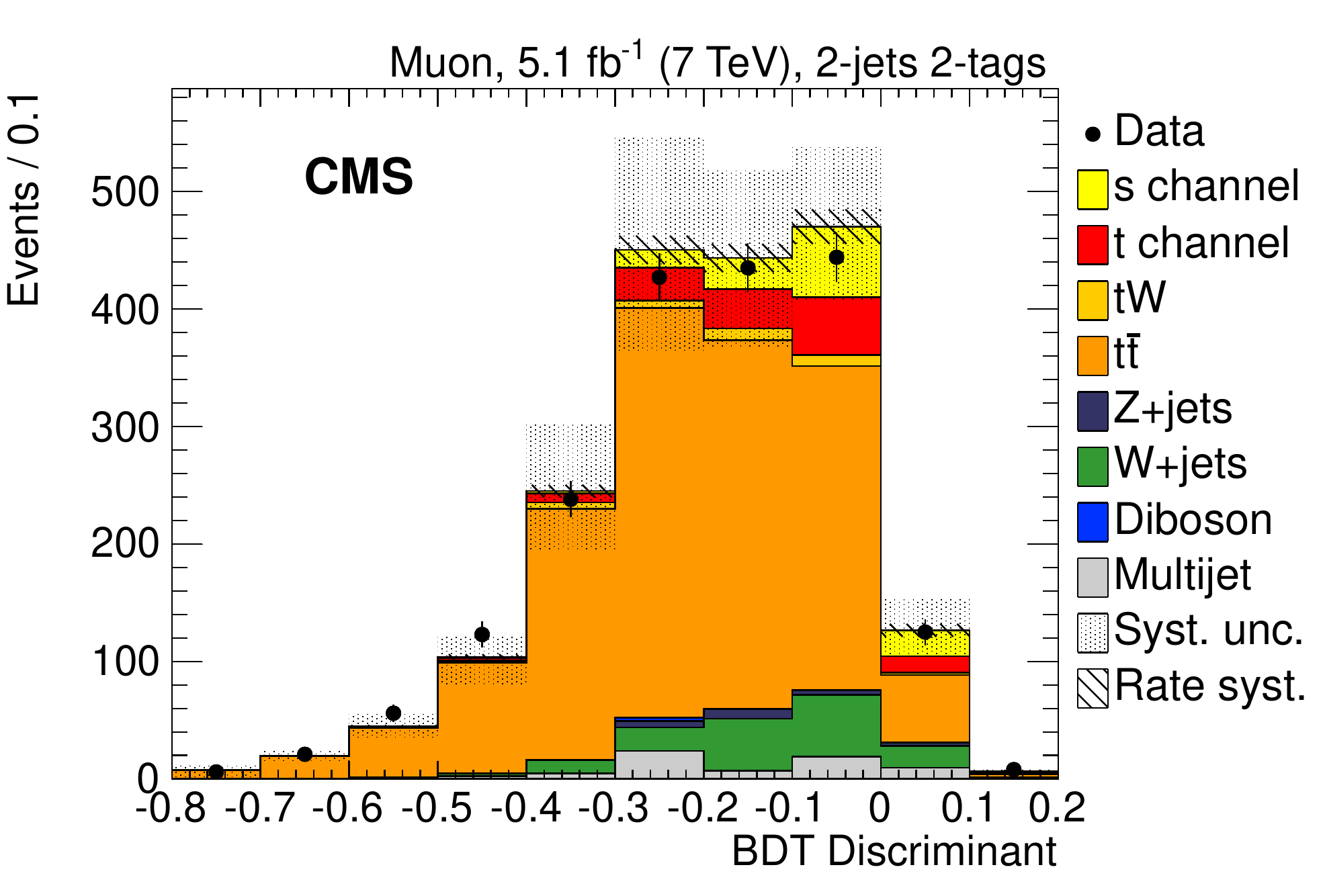} &
    \includegraphics[width=.3\textwidth]{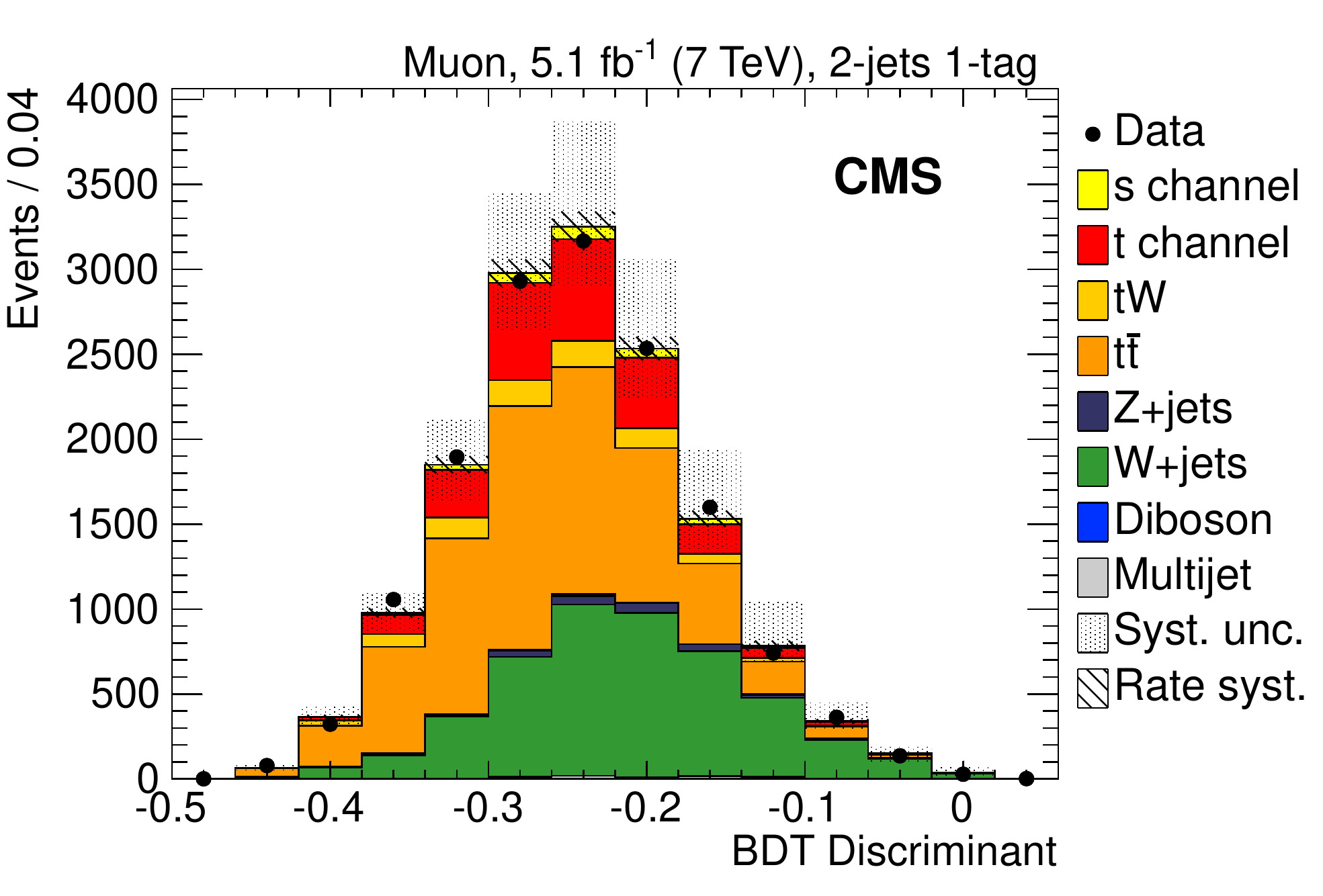} &
    \includegraphics[width=.3\textwidth]{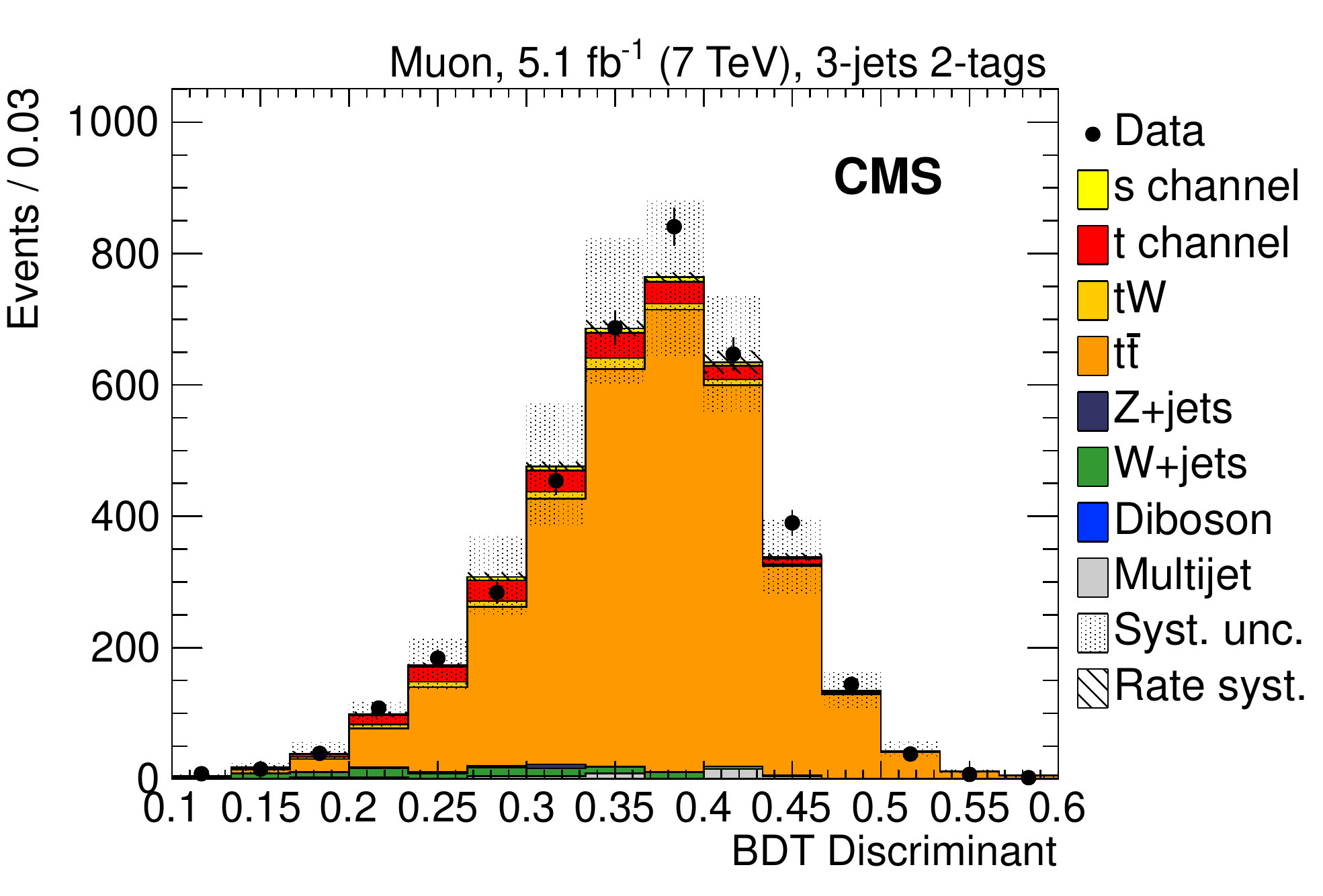}\\
    (a) & (b) & (c) \\
  \end{tabular}
    \caption{Comparison of data with simulation for distributions of the BDT discriminants in the (a) 2-jets 2-tags, (b) 2-jets 1-tag, and (c) 3-jets 2-tags event categories, for the muon channel at \SI{7}{TeV}~\cite{Khachatryan:2016ewo}.
    The simulation is normalized to the combined ($7+\SI{8}{TeV}$) fit results.
    The inner uncertainty bands include the post-fit background rate uncertainties only, the outer ones include the total systematic uncertainty obtained summing in quadrature the individual contributions.}
    \label{fig:cms_tb}
\end{figure}

Measurements of the $s$-channel single top-quark production cross-section with $\SI{5.1}{\ifb}$ of $\sqrt{s}=\SI{7}{\TeV}$ data and $\SI{19.7}{\ifb}$ of $\sqrt{s}=\SI{8}{\TeV}$ data are performed with the CMS detector~\cite{Khachatryan:2016ewo}.
Events with one lepton are selected and classified into a 2j2b signal region, and two control regions, 2j1b ($tq$, $W+$jets), and 3j2b (\ttbar).
Separate BDTs are trained for $\SI{7}{\TeV}$ muon+jets and $\SI{8}{\TeV}$ electron+jets and muon+jets channels in each region.
A combined likelihood fit to the BDT discriminants is used in the signal and control regions, shown in Figure~\ref{fig:cms_tb} for the \SI{7}{TeV} analysis.

The \SI{7}{TeV} and \SI{8}{TeV} results and related predictions from {\sc Hathor} calculations~\cite{Aliev:2010zk} are summarized as follows:

\begin{eqnarray*}
  \sigma_{\text{obs}} (\SI{7}{TeV})     =& 7.1 \pm 8.1\, \text{(stat+syst)}\, \text{pb} \\
  \sigma_{\text{NLO+NNLL}} (\SI{7}{TeV})=& 4.29^{+0.19}_{-0.17} \, \text{pb}\\
  \sigma_{\text{obs}} (\SI{8}{TeV})     =& 13.4 \pm 7.3\, \text{(stat+syst)} \, \text{pb}\\
  \sigma_{\text{NLO+NNLL}} (\SI{8}{TeV})=& 5.24^{+0.22}_{-0.20} \, \text{pb}
\end{eqnarray*}
The observed (expected) signal significance estimated in a combined \SI{7}{TeV} and \SI{8}{TeV} fit is $2.5\sigma\, (1.1\sigma)$.

%ATLAS 8 TeV s-channel
%https://atlas.web.cern.ch/Atlas/GROUPS/PHYSICS/PAPERS/TOPQ-2015-01/
\section{ATLAS $tb$ search at $\sqrt{s}=\SI{8}{\TeV}$}
\begin{figure}[htb]
  \begin{center}
    \begin{tabular}{cc}
      \includegraphics[width=0.35\textwidth]{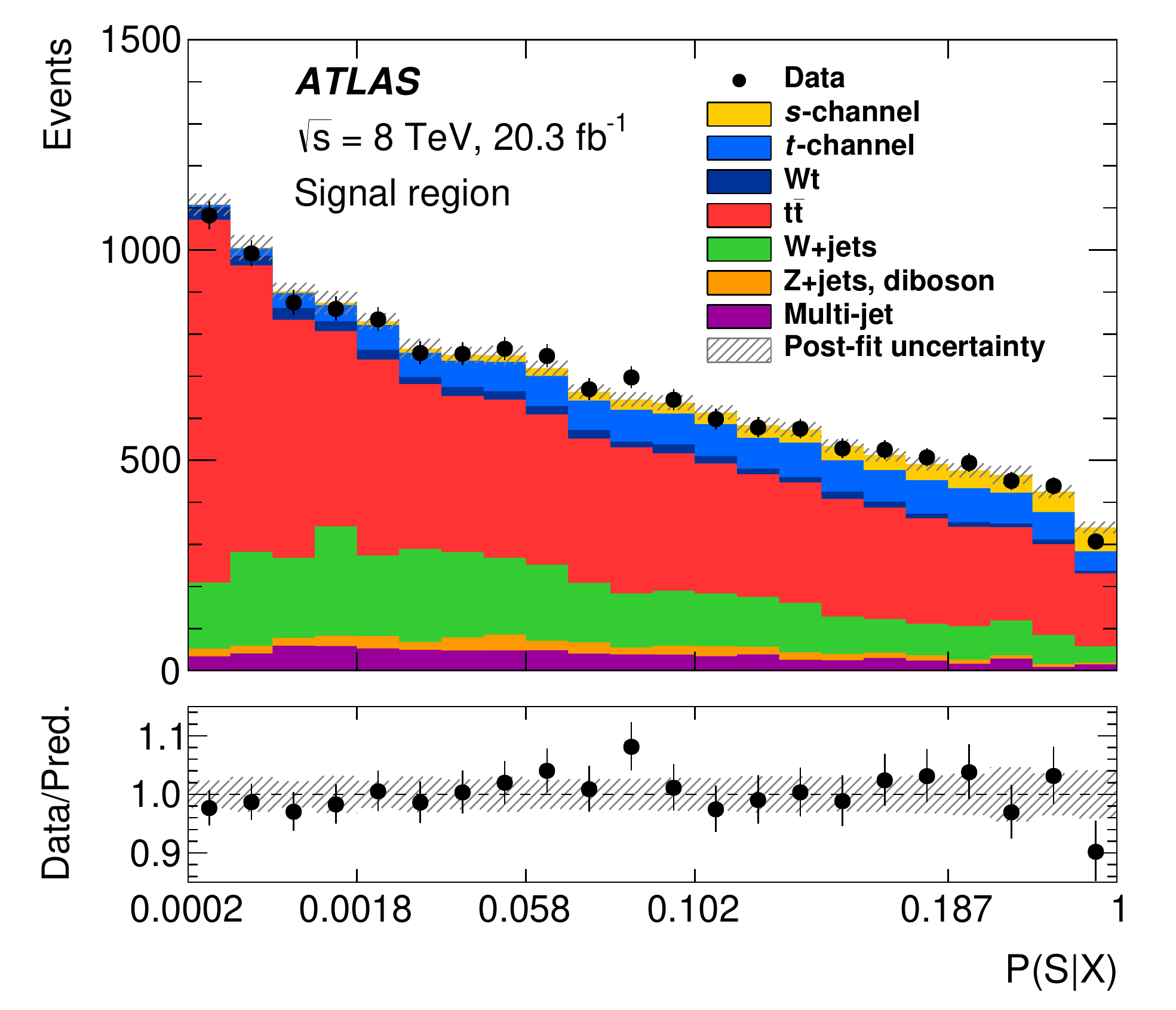} &
      \includegraphics[width=0.35\textwidth]{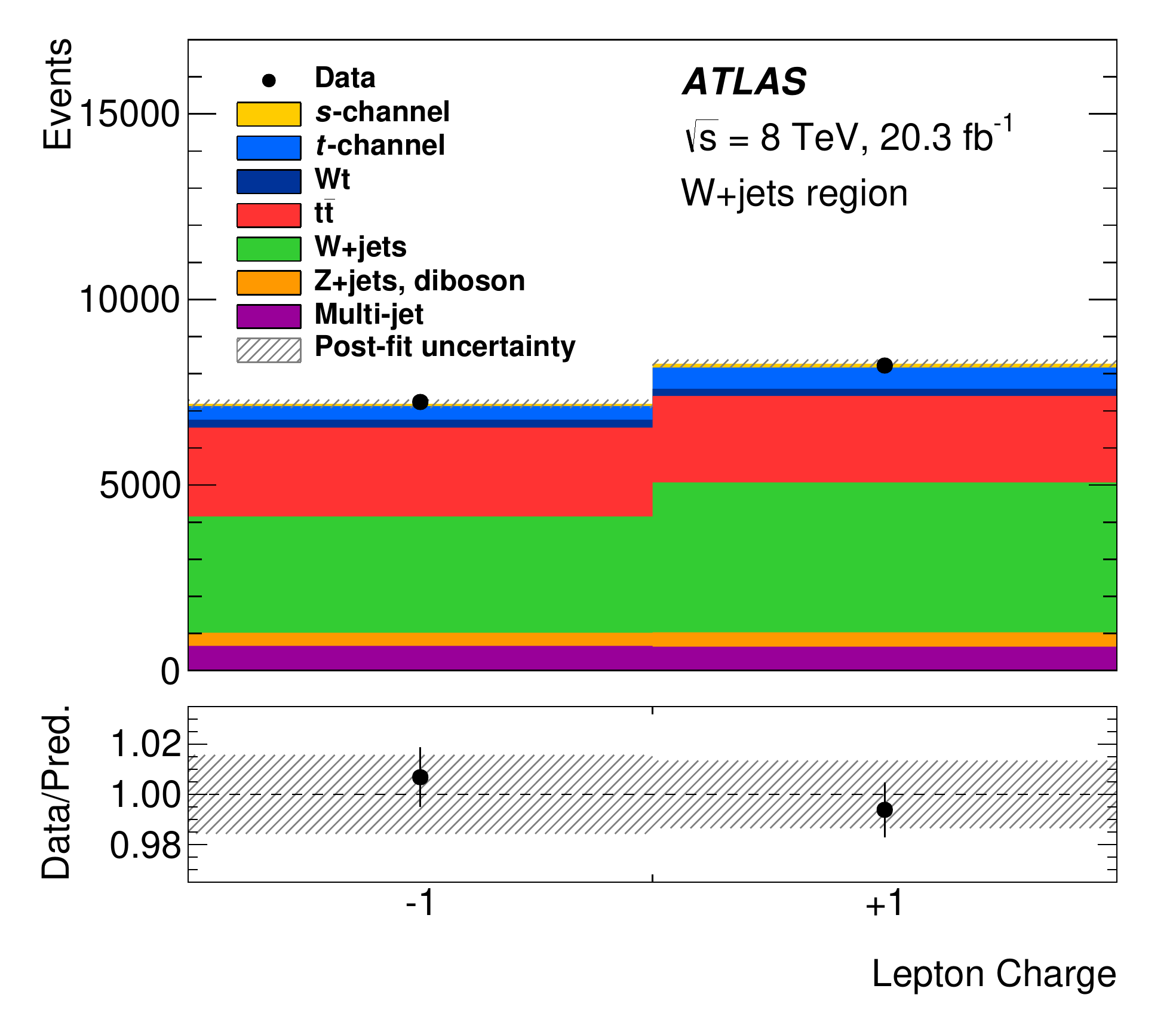} \\
      (a) & (b) \\
    \end{tabular}
  \end{center}
  \caption{Post-fit distribution of the matrix element discriminant in the (a) signal region and (b) $W+$jets control region~\cite{Aad:2015upn}.
  All samples are scaled by the fit result utilizing all fit parameters.
  The hatched bands indicate the total uncertainty of the post-fit result including all correlations.
  The matrix element distributions are made using the optimized binning which is also applied in the signal extraction fit. }
  \label{fig:atlas_tb}
\end{figure}

An analysis of single top-quark production in the $s$-channel is performed by ATLAS with $\SI{20.3}{\ifb}$ of $\sqrt{s}=\SI{8}{\TeV}$ data~\cite{Aad:2015upn}.
Lepton+jets events are selected and classified into two regions: a 2j2b signal region and a 2j1b control region, populated mainly by $t$-channel single top quark and $W+$jets events.
A matrix element method is employed to separate the $tb$ signal from backgrounds.
This method uses information from matrix element calculations for signal and background processes to construct an optimized likelihood ratio discriminant.
A combined likelihood fit to the matrix element discriminant in the signal region and to the lepton charge distribution in the control region is performed, as shown in Figure~\ref{fig:atlas_tb}.
The cross-section measured in the likelihood fit is:

\begin{equation*}
  \sigma_{\text{obs}} = 4.8 \pm 0.8\, \text{(stat)} ^{+1.6}_{-1.3}\,\text{(syst) pb},
\end{equation*}
which is in agreement with the {\sc Hathor} calculation~\cite{Aliev:2010zk}, $\sigma_{\text{NLO+NNLL}} = 5.24^{+0.22}_{-0.20} \,\text{pb}$.
The observed (expected) signal significance is $3.2\sigma\, (3.9\sigma)$, finding evidence for $tb$ production in $pp$ collisions at the LHC.

\section{Summary}
The ATLAS and CMS experiments have produced analyses of single top-quark production in the $s$-channel, via $tW$ production, and via $tZq$ production, establishing cross-section measurement for each process, while observing $tW$ production and finding evidence for $s$-channel production at the LHC.
Future results with Run~2 data collected in 2016 and onward will allow analyses to continue to reduce uncertainties and observe  rarer processes.

\end{document}